\documentclass[twocolumn,showpacs,preprintnumbers,floatfix,letterpaper,prl]{revtex4}
\usepackage{citesort,enumerate}
\usepackage{amsmath,amssymb} 
\usepackage{bm}              
\usepackage{amscd}           
\usepackage{epsfig}          
\usepackage{hhline,multirow} 
\usepackage{dcolumn}         

\newcommand{\eqeqref}[1]{Eq.~\eqref{#1}}
\newcommand{\eqseqref}[1]{Eqs.~\eqref{#1}}
\newcommand{\refref}[1]{Ref.~\cite{#1}}
\newcommand{\refsref}[1]{Refs.~\cite{#1}}

\newcommand{\figref}[1]{Fig.~\ref{#1}}
\newcommand{\tabref}[1]{Table~\ref{#1}}

\newcommand{\beq}{\begin{equation}}
\newcommand{\eeq}{\end{equation}}
\newcommand{\bea}{\begin{eqnarray}}
\newcommand{\beas}{\begin{eqnarray*}}
\newcommand{\beau}[1]{\begin{equation} \label{#1} \begin{array}{rcl}}
\newcommand{\eea}{\end{eqnarray}}
\newcommand{\eeas}{\end{eqnarray*}}
\newcommand{\eeau}{\end{array} \end{equation}}
\newcommand{\bay}{\begin{array}}
\newcommand{\eay}{\end{array}}
\newcommand{\bals}{\begin{align*}}
\newcommand{\eals}{\end{align*}}

\newcommand{\ds}{\displaystyle}
\newcommand{\fs}{\footnotesize}

\newcommand{\ra}{{\rightarrow}}

\newcommand{\vev}[1]{\langle #1 \rangle}

\newcommand{\In}[1]{ \bigg| _{#1} }

\begin{document}



\title{Cronin effect vs. geometrical shadowing
in d+Au collisions at RHIC}

\author{A.Accardi}%
\author{M.Gyulassy}%
\affiliation{%
Columbia University, Department of Physics\\
538 West 120th Street, New York, NY 10027, USA
}
\date{August 9th, 2003}

\begin{abstract}
Multiple initial state parton interactions in $p(d)+Au$ collisions
are calculated in a Glauber-Eikonal formalism. The convolution of
perturbative QCD parton-nucleon cross sections predicts naturally the
competing pattern of low-$p_T$ suppression due geometrical shadowing,
and a moderate-$p_T$ Cronin enhancement of hadron spectra. The formal 
equivalence to recent classical Yang-Mills calculations is demonstrated, 
but our approach is shown to be more general in the large $x>0.01$ domain
because it automatically incorporates the finite kinematic constraints of both 
quark and gluon processes in the fragmentation regions, and
accounts for the observed spectra in elementary
$pp\rightarrow \pi X$ processes in the RHIC energy range,
$\sqrt{s}\sim 20-200$ GeV.   The Glauber-Eikonal formalism can be used
as a baseline to extract the magnitude of dynamical shadowing effects
from the experimental data at differente centralities and pseudo-rapidities.
\end{abstract}


\maketitle


It is well known that in proton ($p$), or deuteron ($d$), 
reactions involving heavy nuclei ($A\sim 200$) at 
$\sqrt{s}<40$ AGeV, the moderate transverse momentum
($p_T\sim 2-6$ GeV) spectra are enhanced relative to linear extrapolation from 
$p+p$ reactions. This Cronin effect
\cite{Cronin75,Antreasyan79,Straub92} is generally attributed to multiple
scattering of projectile partons propagating through the target nucleus. 
The data  can be well accounted for phenomenologically by adding a
random Gaussian  
transverse kick $\delta k_T^2\propto \mu^2 L/\lambda$ to the
projectile partons prior to
hadronization \cite{ktbroadening1,ktbroadening2,Vitev03,ktbroadening3}. 
Here  $\lambda$ is
the parton mean free path in the nucleus, $L\propto A^{1/3}$ is the 
 average path length, and $\mu$ is a typical screening mass in ground state  nuclei.
These models naturally predict a slowly decreasing Cronin effect with 
increasing energy  which has only recently been possible
to test at $\sqrt{s}=200$ AGeV 
at the Relativistic Heavy Ion Collider (RHIC).

Interest in the Cronin effect has been revived,
due to the developement of a new formulation of the physics
based on the concept of gluon saturation
and classical Yang-Mills field models \cite{newsatmodels,CGCreviews}.
In addition, a radical 
possibility was proposed  in \refref{CGCandsuppression},
that nonlinear gluon saturation may in fact strongly suppress 
moderate-$p_T$ spectra at RHIC. 
If such a deep gluon shadowing in this kinematic range were true, 
then an anti-Cronin suppression should have been observed in the large
$x=2 p_T/\sqrt{s}\approx 0.01-0.1$ and moderate $1<Q=p_T<10$ GeV scale
range accessible at RHIC.
Four experiments at RHIC \cite{PHENIX,STAR,PHOBOS,BRAHMS} found 
independently that the nuclear modification factor, $R_{dA}(p_T)= 2
d\sigma_{dA}/Ad\sigma_{pp}$, 
 was consistent with a positive Cronin enhancement of hadrons with
$2<p_T<5$ GeV. The magnitude of the enhancement is somewhat smaller than
predicted for the $\pi^0$, as reviewed in \cite{Accardi02}, 
but no evidence of strong shadowing was reported.

After the release of the RHIC data, the
saturation model  predictions of suppression 
were revised in \refref{BKW03,KKT03,Albacete03}.
In \refref{BKW03} Cronin enhancement with no high-$p_T$ suppression
was shown to be a generic feature of saturation models.
In \refref{KKT03} another version of saturation dynamics with Cronin
enhancement coupled with the high-$p_T$ suppression of 
\refref{CGCandsuppression} was discussed. 
In \refref{Albacete03}, the Cronin enhancement at $y=0$ was predicted to be 
progressively negated  by non linear QCD evolution at smaller nuclear
$x$, and therefore a gluon shadowing suppression is 
predicted at higher rapidities. 

Different approaches to the calculation of the Cronin effect 
can be formulated in infinite momentum and target frames.
In the traditional Glauber-Eikonal (GE)  approach \cite{GEmodels,AT01b,GLV02}, 
sequential multiple partonic collisions 
in the target frame are computed. This leads to transverse diffusion and unitarity
is naturally
preserved. The low-$p_T$ spectra are suppressed by unitarity 
to compensate for the moderate $p_T$ Cronin enhancement. 
This is what we call ``geometrical shadowing'', as it is driven
by the geometry of the collision.
No high-$p_T$ shadowing is predicted in this approach. 

In applications, the GE series 
has been directly evaluated thusfar only up to the three-scattering term and
for $\sqrt s \leq 40$ GeV \cite{GEmodels,Straub92}. 
Numerically more convenient 
approximated GE models \cite{ktbroadening1,ktbroadening2,Vitev03,ktbroadening3}
have been proposed. They modify the pQCD rates 
through the inclusion of a nuclear broadened intrinsic $k_T$, instead  
of evaluating the full GE series.
Phenomenologically, it is well known \cite{Field89,Owens87}
 that intrinsic $k_T\sim 1$ GeV must be introduced
to correct collinear factorized pQCD
predictions to account for  $p+p$ data at moderate $p_T<5$ GeV.
The approximated GE models simply extend that idea by  adding a random kick 
$\delta k_T^2$ to the intrinsic $k_T^2$. 
One drawback of such approaches is that 
 a non-trivial $p_T$ or collision number 
dependence of the effective nuclear transport coefficient
$\mu^2/\lambda\sim 0.05$ GeV$^2$/fm \cite{ktbroadening2}
must also be introduced to account for the actual Cronin data.
While a logarithmic $p_T$ dependence of $\delta k_T^2(p_T) $ is expected
for partons undergoing multiple Yukawa screened interactions \cite{GLV02},
the functional form of that $p_T$ dependence is usually
adjusted to fit the Cronin data at one energy.
A further drawback of such approximated GE models is
that the unitarity constraints built into GE are ignored and
hence the unitarity shadow and Cronin are treated as two seperate phenomena.

The more recent approaches \cite{DM02,DJ02,GJ03,JNV03} to  
the Cronin effect in the infinite momentum frame are based on 
the Mclerran-Venugopalan (MV) model of the nuclear wave funtion in
classical Yang-Mills theory \cite{MVmodel}. The  general equivalence of GE and MV formulations for transverse diffusion
was discussed in \cite{KM98,Kovchegovetal01} in the context of
gluon dominated 
small $x\ll 1$ kinematics.
In these approaches, the
nucleus is approximated by a Weisz\"acker-Williams 
gluon field with non-linearities approximated  semi-analytically or
computed numerically \cite{KNV,JNV03}. 
The non-linear gluon interactions lead to transverse diffusion and hence Cronin enhancement
of  nuclear partons prior to the scattering.
The essential scale in this approach is a gluon saturation scale
$Q_s=Q_s(y,\sqrt{s},A)$, with $y$ the rapidity of the produced gluon.
One of the  advantages of the MV approach is that unitarity
is at least 
approximately enforced through the conservation of the number of virtual gluons
in the transverse diffusion. Therefore these models predict a definite 
 anti-Cronin suppression
below some scale $\propto Q_s$. On the other hand, a disadvantage in
present formulations of the MV model 
is that they ignore finite-energy kinematics of 
valence- and sea quark-induced processes and the non-asymptotic
 large $x>0.01$ features of gluon structure, where the classical approximation
is unrealistic. A major disadvantage of MV models is that
they cannot account for  the elementary $p+p$ transverse spectrum,
that forms the denominator of the $R_{pA}$ nuclear modification
factor. Neither can the models reproduce 
the absolute normalization of the spectra in $pA$ collisions without extra
phenomenological assumptions.

 In this letter, we compute directly 
the GE series via numerical convolution of elementary
parton-nucleon processes.  An advantage over approximated GE models 
\cite{ktbroadening1,ktbroadening2,Vitev03,ktbroadening3}
is that our approach automatically conserves unitarity in
the geometric 
optics sense of GE theory. In addition, we do not 
introduce extra phenomenological $p_T$-dependent nuclear broadening of
the intrinsic $k_T$, since the GE series predicts the functional form
of the Cronin enhancement 
based on the calculated non-asymptotic pQCD parton-nucleon cross sections. 
An  advantage of our GE approach over MV model applications is that we
autmatically include the finite kinematic large-$x$ feautures of both
quark and gluon processes.  
Perhaps the most  important advantage over the MV approaches is that our formulation
is directly constrained to reproduce the  absolute normalized spectra
in $p+p$ collisions. Therefore the GE 
approach presented below calculates consistently both $p+p$ and
$p+A$ collisions at the finite energies accessible at RHIC. 

Beside the geometrical quark and gluon shadowing, which is automatically
included in GE models, at low enough $x$ one expects
genuine dynamical shadowing due to non-linear gluon interactions, as
described in saturation models.  
Both kind of shadowing are present in the data, but it is not possible a
priori to tell in which proprotion. The Glauber-Eikonal formalism can
then be used as a baseline to extract the magnitude of dynamical
shadowing effects from the experimental data.


\section{Parton-nucleus collisions in the Glauber-Eikonal model}

The GE expression for a parton nucleus scattering
\cite{AT01b,GLV02} is:
\begin{align}
  \frac{d\sigma^{\,iA}}{d^2p_Tdyd^2b}
     = & \sum_{n=1}^{\infty} \frac{1}{n!} \int d^2b \, d^2k_1 \cdots d^2k_n
     \nonumber\\*
  & \times \frac{d\sigma^{\,iN}}{d^2k_1} T_A(b) 
     \times \dots \times  
     \frac{d\sigma^{\,iN}}{d^2k_n} T_A(b)
     \nonumber\\*
  & \times e^{\, - \sigma^{\,iN}(p_0) T_A(b)} \,\, 
      \delta\big(\sum _{i=1,n} {\vec k}_i - {\vec p_T}\big) \ ,
 \label{dWdp}
\end{align}
where $T_A(b)$ is the target nucleus thickness function at impact
parameter $b$. The differential and integrated
parton-nucleon cross sections,
$d\sigma^{\,iN}/{d^2k}$ and  $\sigma^{\,iN}(p_0) = \int {d\sigma^{\,iN}}$,
 are  computed in pQCD as discussed below, see
\eqeqref{avflux}-\eqref{iNxsec}. The integrated
parton-nucleon cross section depends on an infrared
scale $p_0$, which is determined by fitting $p+p$ data.
The exponential factor in \eqeqref{dWdp} represents the probability
that the parton suffered no semihard scatterings after the $n$-th one. 
In such a way, unitarity is explicitly implemented at the
nuclear level, as discussed in Ref.~\cite{AT01b}.
The sum over $n$ starts from $n=1$ because we are interested in partons
which are put on-shell by the interaction and later on hadronize.
The sum over $n$ may be performed in Fourier space.
The result reads:
\begin{subequations}
\begin{align}
  \frac{d \sigma^{\,iA}}{d^2p_Tdyd^2b} = \int \frac{d^2r_T}{4\pi^2}
    e^{\,-i \vec{k}_T \cdot \vec r_T} S^{\,iA}(r_T;p_0) \ ,
 \label{spechard}
\end{align}
where, suppressing the dependence of the quantities in the r.h.s. on
$y$,
\begin{align}
    S^{\,iA}(r_T;p_0) =  
       e^{\,-\,\widetilde\sigma^{\,iN}(r_T,p_0)\,T_A(b)} 
         - e^{\,-\sigma^{\,iN}(p_0) T_A(b)}
 \label{Shard}
\end{align}
and
\begin{align}
        \widetilde\sigma^{\,iN}(r_T;p_0) = \int d^2 k
        \left[ 1 - e^{\,-i \vec k \cdot \vec r_T} \,\right]
        \frac{d\sigma^{\,iN}}{d^2kdy_i} \ .
 \label{sighard}
\end{align}
\end{subequations}
It is possible to show \cite{AT01b}
that, in the high-$p_T$ limit, \eqeqref{pAcoll}
reduces to the usual single scattering approximation for
parton-nucleus scattering:
\begin{equation}
\begin{CD}
  \ds \frac{d\sigma}{dp_T^2 dy d^2b}^{\hspace{-0.5cm}iA\rightarrow iX}
    \hspace*{-.6cm}
    @>>{p_T\ra\infty}> 
    \ds T_A(b) \frac{d\sigma}{dp_T^2 dy}^{\hspace{-0.3cm}iN\rightarrow
    iX} \ .
\end{CD}
\vspace*{.0cm}
 \label{highptlim}
\end{equation}
As $p_T \ra 0$ unitarity corrections switch on, suppressing the
integrated parton yield \cite{AT01a}, 
and inducing a random walk of
the parton in $p_T$ space, thus redistributing the partons to higher
$p_T$ compared to the single scattering approximation \cite{AT01b}.
This is how the multiple scattering mechanism of \eqeqref{dWdp} induces 
``geometrical'' shadowing at low $p_T$, and 
Cronin enhancement of the transverse spectrum at moderate $p_T$,
respectively. Note also that the accumulation of $k_T$ kicks in the
multiple scattering process is computed in the model, not input as
a Gaussian folding as in approximated GE models.
The full expression for the $p_T$ spectrum, \eqseqref{pAcoll} and
\eqref{spechard}-\eqref{sighard}, interpolates naturally 
between the geometrical shadowed low-$p_T$ and the Cronin enhanced
moderate-$p_T$ regions.

Note that $\widetilde\sigma^{\,iN}(r_T) \propto r_T^2$ as
$r_T\rightarrow 0$ and $\widetilde\sigma^{\,iN}(r_T) \rightarrow
\sigma^{\,iN}$ as $r_T\rightarrow \infty$. This suggests the
interpretation of $\widetilde\sigma^{\,iN}(r_T)$ as a  
dipole-nucleon ``hard''  cross section. This dipole is of
mathematical origin, and comes from the square of the scattering
amplitude written in the Fourier variable $r_T$, which represents the
transverse size of the dipole.
Then, we can interpret $S^{\,iA}$, \eqeqref{Shard}, as the
dipole-nucleus ``hard'' cross section, which clearly incorporates
Glauber-Gribov multiple scatterings of the colour dipole.
No other nuclear effects on PDF's are included beside
multiple scatterings.

The interpretation of the interaction in terms of multiple scatterings
of a dipole allows to relate this approach to other multiscattering 
formalisms such as \refref{INSZZ02,KNST02} and the saturation
computations of \refref{DJ02,GJ03,Kovchegovetal01,Albacete03}. 
A first step in building such a dictionary was taken in \refref{AG03},
where it was shown that the dipole cross sections of 
\eqseqref{Shard}-\eqref{sighard} are equivalent, in a suitable
kinematic region, to the dipole cross section
considered in the MV model of \refref{DJ02,GJ03}. 
The main 
difference is in the input
parton-nucleon cross section in $p_T$-space. 
In our case, as we will discuss in the
next sections, it is computed in pQCD, including full kinematics and
interactions of the incoming parton with both quarks and gluons.
Moreover, its energy and rapidity dependence are controlled by the
DGLAP evolution of the parton distribution functions of the target
nucleon. In the MV model, it is approximated using asymptotic
kinematics for gluon targets only.  
  Both models consider in $S^{iA}$ only inelastic 
  dipole-nucleus scatterings. They neglect diffrective dipole-nucleus
  interactions, which however modify the $p_T$-spectrum only at
  the lowest tarnsverse momenta \cite{KolyaYuri}.

Beside the geometrical quark and gluon shadowing, which is automatically
included in GE models, at low enough $x$ one expects
genuine dynamical shadowing due to non-linear gluon interactions as
described in the saturation models. However, it is difficult to
disentangle these two sources of shadowing and suppression.
The distinction between the two is however of
fundamental interest as has already been emphasized
in the context of $e+p$ DIS 
 HERA by A.~Caldwell \cite{Caldwelpc}. Most theoretical interest is not
in the ubiquitous geometrical shadowing and unitarity corrections, but
in the onset of genuine nonlinear QCD physics 
\cite{SatatHERA}. 
Moreover, saturation models cannot predict as yet the upper bound on  $x$
below which non linear effects set in.
In order to help recognize possible  novel  nonlinear regimes
it is essential to be able to
calculate the  baseline spectra 
isolating the unitarity and geometrical shadowing alone. 
The parton level GE model discussed below provides such a baseline.


\section{Inclusive minijet and hadron production in $pp$ collisions}

Let's consider a $pp'$ collision, where $p$ and $p'$ stand for a
proton ($p$), a deuteron ($d$), or a nucleon ($N$). 
In leading order pQCD, the inclusive cross section for production of a
parton of flavour $i = g,q,\bar q$ ($q=u,d,s,\dots$)
with transverse momentum $p_T$ and rapidity $y$
\cite{EH03} may be written as a sum of 
contributions to the cross section coming from projectile ($p$) partons 
and from target ($p'$) partons:
\begin{eqnarray}
  \frac{d\sigma}{dp_T^2 dy}^{\hspace{-0.3cm}pp'\rightarrow iX}
    \hspace{-0.7cm} &=& \Bigg\{
    \vev{xf_{i/p}}_{y_i,p_T} \, \frac{d\sigma^{\,ip'}}{dy_i d^2p_T} \In{y_i=y}
    \nonumber \\*
  && + \vev{xf_{i/p'}}_{y_i,p_T} \, 
    \frac{d\sigma^{\,ip}}{dy_i d^2p_T}\In{y_i=-y} \Bigg\} \ .
 \label{ppcoll_pt}
\end{eqnarray}
Here we considered only elastic parton-parton subprocesses,
which contribute to more than 98\% of the cross section at midrapidity
\cite{SEC89}.
In \eqeqref{ppcoll_pt}, 
\begin{widetext}
\begin{subequations}
\begin{eqnarray}
  \vev{xf_{i/p}}_{y_i,p_T} &=& \ds K 
      \sum_j \frac{1}{1+\delta_{ij}}  
      \int dy_2 x_1f_{i/p}(x_1,Q_p^2) 
      \frac{d\hat\sigma}{d\hat t}^{ij} \hspace{-0.2cm} 
      (\hat s,\hat t,\hat u) \, x_2f_{j/p'}(x_2,Q_p^2) 
    \Bigg/ 
      \frac{d\sigma^{ip'}}{d^2p_T dy_i}
 \label{avflux}
    \\*
  \frac{d\sigma^{ip'}}{d^2p_T dy_i} &=& K 
    \sum_j \frac{1}{1+\delta_{ij}} 
    \int dy_2 \frac{d\hat\sigma}{d\hat t}^{ij} \hspace{-0.2cm} 
    (\hat s,\hat t,\hat u) \, x_2f_{j/p'}(x_2,Q_p^2) 
 \label{iNxsec}
\end{eqnarray}
\end{subequations}
\end{widetext}
are interpreted, respectively, as the average flux of incoming partons
of flavour $i$ from the hadron $p$, and the cross section for the
parton-hadron scattering.
The rapidities of the $i$ and $j$ partons in the final state 
are labelled by $y_i$ and $y_2$. 
Infrared regularization is performed by adding a small mass to the
gluon propagator and defining 
$m_{T}=\sqrt{p_T^2+p_0^2}$. The fractional momenta of the colliding partons
$i$ and $j$ are $x_{1,2}=\frac{m_T}{\sqrt s}({\rm e}^{\pm y_i}+{\rm e}^{\pm
y_2})$, i.e. the incoming partons are collinear with the beams. 
The integration region for $y_2$ is $-\log(\sqrt s/m_T-{\rm e}^{-y_i})\le 
y_2\le \log(\sqrt s/m_T-{\rm e}^{y_i})$.
The summation runs over parton flavours $j = g,q,\bar q$.
The partonic Mandelstam variables are
\begin{eqnarray*}
  \hat t &=& - m_T^2 (1+e^{-y_i+y_2}) \\
  \hat u &=& - m_T^2 (1+e^{ y_i-y_2}) \\
  \hat s &=& - \hat t - \hat u = x_1 x_2 s\ .
\end{eqnarray*}
For the parton distributions we use the CTEQ5 parametrization
at leading order \cite{CTEQ5}. The choice of the factorization scale
$Q_p$ is  discussed later. 
The cross sections $d\hat\sigma^{ij}/d\hat t$ of the
$ij\ra ij$ elastic partonic subprocesses can be found, e.g., 
in \cite{Field89}. They are proportional to $\alpha_{\rm s}(\mu^2)$,
computed as in \cite{EH03},  at a scale $\mu=Q_p$ 
The factor $K$ in \eqeqref{ppcoll_pt} is introduced in order to
account for next-to-leading order (NLO) corrections \cite{AFGKW00}, 
and is in general $\sqrt s$ and scale dependent \cite{EH03,Kfits}.

Inclusive hadron production through independent fragmentation of
the parton $i$ into a hadron $h$, is computed as a convolution of the
partonic cross section \eqref{ppcoll_pt} with a fragmentation function
$D_{i\rightarrow h}(z,Q_h^2)$: 
\begin{eqnarray}
  \frac{d\sigma}{dq_T^2dy_h}^{\hspace{-0.35cm}pp'\rightarrow hX}
    \hspace{-0.6cm} &=& 
    \frac{d\sigma^{\,pp'\rightarrow iX}}{dp_T^2 dy_i}
    \otimes D_{i\rightarrow h}(z,Q_h^2) \ ,
 \label{pptohadron}
\end{eqnarray}
where $q_T$ is  the transverse momentum of the hadron $h$, $y_h$
its rapidity, and $z$ the light-cone fractional momentum of the hadron
and of its parent parton $i$. 
For details, see Eqs.~(8)-(11) of \refref{EH03}. In this
letter, we use LO Kniehl-Kramer-P\"otter fragmentation functions 
\cite{KKP} and set the fragmentation scale $Q_h=Q_p$.  

\begin{figure}[t]
\begin{center}
\parbox[c]{8.5cm}{\epsfig{figure=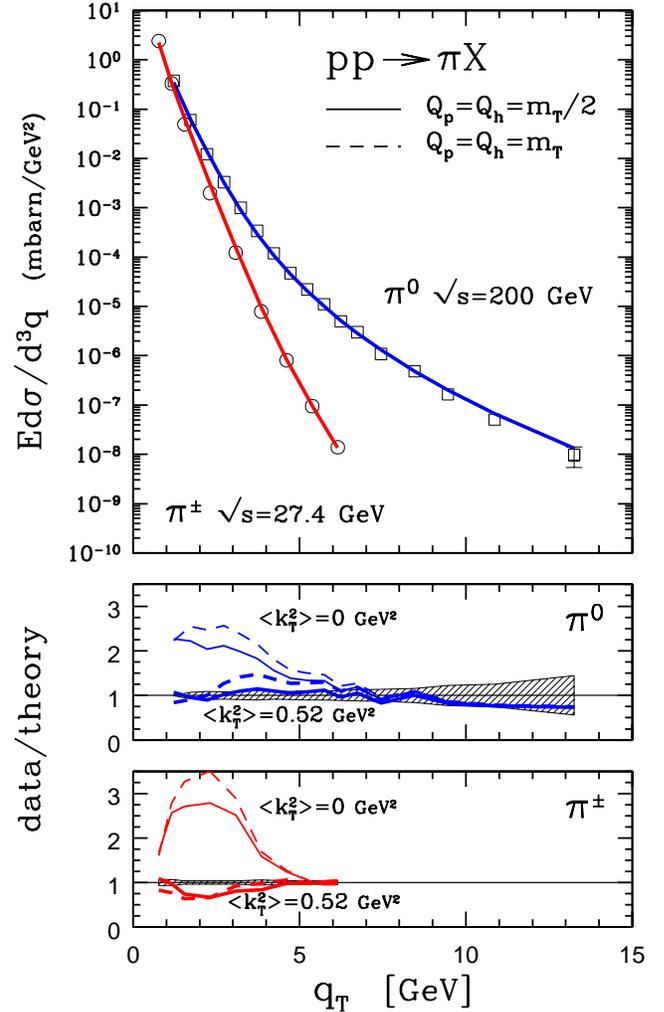,width=8.5cm
,clip=,bbllx=215pt,bblly=150pt,bburx=575pt,bbury=720pt}}
\vskip-.2cm
\caption{\footnotesize
{\it Top panel:} pion transverse momentum spectrum at $\sqrt s = 27.4$
GeV and $\sqrt s = 200$ GeV. Solid lines are LO pQCD computations
according to \eqeqref{ppcoll_pt}, with $\vev{k_T^2}=0.52$ GeV$^2$ and
$Q_p=Q_h=m_T/2$. The regulator $p_0$ and the $K$-factor are given in
\tabref{table:p0Kfits}. {\it Bottom panels:} the data to theory ratio for
different choices of parameters. Solid lines are for $Q_p=Q_h=m_T/2$,
dashed lines $Q_p=Q_h=m_T$. The pair of thin lines is computed with no
$k_T$ smearing, $\vev{k_T^2}=0$ GeV$^2$, and the pair of thick lines is
computed with $\vev{k_T^2}=0.52$ GeV$^2$. In the $\pi^0$ case
\cite{Adler03pp}, the dashed area shows the relative statistical and
point-to-point systematic error added in quadrature, and does not
include the systematic uncertainty of 9.6\%, on the absolute
normalization of the spectrum. In the $\pi^\pm$ case
\cite{Antreasyan79}, the shaded area
includes statistical error only, without a systematic uncertainty of
20\% on the absolute normalization of the spectrum. 
}
\label{fig:data-theory}
\end{center}
\vskip-.4cm
\end{figure}

In the computation of the $pp'$ cross section \eqref{pptohadron}, we
have two free parameters, $p_0$ and $K$, and a somewhat arbitrary
choice of the factorization, renormalization and fragmentation scales.
Our strategy is to compare two choices for those scales, namely
$Q_p=Q_h=m_T/2$ and $Q_p=Q_h=m_T$, and then fit $p_0$, $K$ 
to hadron production data in $pp$ collisions at the
energy of interest. We analyze here $\pi^\pm$ production at
$\sqrt s = 27.4$ GeV \cite{Antreasyan79}, and $\pi^0$ production at
$\sqrt s = 200$ GeV \cite{Adler03pp}.  
For the $K$-factor we perform a $\chi^2$ fit to the high-$p_T$ tail of
the data, following the procedure described in \refref{EH03}.
The fit of $p_0$ is performed by requiring that the computed spectrum 
does not exceed the experimental data at low $p_T\lesssim 1$ GeV. At
$\sqrt s =200$ GeV, this fit is difficult because data exist for 
$q_T>1.2$ GeV only, so we used also data on charged hadron production 
\cite{Adams03}. The resulting data/theory ratio is plotted with thin
lines in the 
bottom panel of \figref{fig:data-theory}, and the extracted parameters are
listed in \tabref{table:p0Kfits}. Note that the $K$-factor is strongly
correlated to the choice of scale, while $p_0$ is more stable. 
Both of them depend on $\sqrt s$.

As in \refref{EH03}, we obtain a satisfactory description of data 
for $q_T\gtrsim 5$ GeV over a broad range $\sqrt s$, 
but the curvature of the hadron spectrum is overpredicted in the
$q_T=1-5$ GeV range. 
As it is well known \cite{Owens87,Field89}, this can be corrected by
considering an intrisic transverse momentum $k_T$ for the colliding
partons \cite{Kfits}. 
There exists many ways of implementing it phenomenologically,
and we choose for simplicity a $k_T$ smearing of the
cross section to approximate this effect. We introduce then 
unintegrated parton distributions
\begin{equation*}
  \tilde f_i (x,\vec k_T,Q_p^2) =  
    \frac{e^{-k_T^2/\vev{k_T^2}}}{\pi \vev{k_T^2}} f_i (x,Q_p^2) \ ,
\end{equation*}
where the width $\vev{k_T^2}$ of the Gaussian enters as a
phenomenological parameter, and convolute over $d^2 k_{1T}$ and $d^2
k_{2T}$ in \eqseqref{avflux}-\eqref{iNxsec}.

We found that a fixed $\vev{k_T^2}=0.52$ GeV 
leads to a dramatic improvement 
in the computation of the trasverse spectra, which now agrees 
with data at the $\pm$40\% level. 
The quality of our pQCD computation including 
intrinsic $k_T$ is shown in \figref{fig:data-theory}, and the
extracted $K$ is reported in \tabref{table:p0Kfits}. Without intrinsic
$k_T$ it is not possible to fit the value of $p_0$, due to the
steepness of the data/theory ratio. The fit of the $K$-factor in this
case was made with the $p_0$ determined using the intrinsic $k_T$.
The optimal choice of scale is found to be $Q_p=Q_h=m_T/2$ at both
energies, as the value of the $K$-factor is the closest to 1.

\begin{table}[tbp]
\begin{ruledtabular}
\vskip.2cm
\begin{footnotesize}
\begin{tabular}{cc|c|c}
 $\vev{k_T^2}$ \hspace*{-.2cm}& $Q_p=Q_h$ & $\sqrt s=27.4$ {\fs GeV} & $\sqrt s=200$ {\fs GeV}
 \\\hline 
 \multirow{4}{1.3cm}{\centerline{0.52 GeV$^2$}} \hspace*{-.2cm}
   &\multirow{2}{0.9cm}{\centerline{$m_T/2$}}
          & $p_0=0.70 \pm 0.1$ GeV & $p_0=1.0 \pm 0.1$ GeV\\
         && $K=1.07 \pm 0.02$ & $K=0.99 \pm 0.03$ \\
   &\multirow{2}{0.9cm}{\centerline{$m_T$}}
          & $p_0=0.85 \pm 0.1$ GeV& $p_0=1.2 \pm 0.1$ GeV\\
         && $K=4.01 \pm 0.08$ & $K=2.04 \pm 0.12$ 
 \\ \hline
 \multirow{4}{1.3cm}{\centerline{0 GeV$^2$}} \hspace*{-.2cm}
   &\multirow{2}{0.9cm}{\centerline{$m_T/2$}}
          & $p_0=$ ---  & $p_0=$ --- \\
         && $K=3.96 \pm 0.11$ & $K=1.04 \pm 0.06$ \\
   &\multirow{2}{0.9cm}{\centerline{$m_T$}}
          & $p_0$= ---  & $p_0 =$ ---\\
         && $K=13.4 \pm 0.4$ & $K=2.04 \pm 0.12$ 
\end{tabular}
\end{footnotesize}
\end{ruledtabular}
\caption{Fitted values of $p_0$ and the $K$-factor for $\pi^\pm$ and $\pi^0$
  production  in $pp$ collisions at $\sqrt s=27.4$ GeV and  $\sqrt
  s=200$ GeV, respectively. We quoted the fit uncertainties only. The
  systematic uncertainty in the absolute normalization of experimental
  data (20\% and 9.6\%, respectively), which affect the determination
  of the K-factor are not included. The fit of the $K$-factor in the
  case of zero intrinsic momentum was made with the $p_0$ determined 
  using the $\vev{k_T^2}=0.52$ GeV$^2$.}
\label{table:p0Kfits}
\end{table}

Let us comment briefly on the physical meaning of the infrared
regulator $p_0$. The divergence of the pQCD cross section
for minijet production, indicates clearly a break-down of unitarity at
low $p_T$. A phenomenological way of restoring it, is to tame the
divergence of $d\hat\sigma^{ij}/d\hat t$ by adding a small mass
regulator, $p_0$, to the exchanged transverse momentm $p_T$. 
Seen in this light, $p_0$ represents the scale at which higher order
parton processes enter into the game, beside the single scatterings
considered in \eqseqref{avflux}-\eqref{iNxsec}. As the center of mass energy is
increased, the partons probe the nucleon at smaller $x$, so
that the density of target parton increases, and one should expect
unitarity effects to arise at larger scales. The slight increase of
the fitted $p_0=p_0(\sqrt s)$ with energy is indeed a consistent check
of this picture.
For the same reason, and since quarks interact more weakly than gluons,
one might expect also the unitarity corrections for quark-nucleon
scattering to arise at $p_{0|\text{quarks}} \leq
p_{0|\text{gluons}}$. However, to check this relationship, we would
need at least data on $K^\pm$ production in a $p_T$ range of $1-6$ GeV.
Here we simply set
 $p_{0|\text{quarks}} = p_{0|\text{gluons}}=p_0$.

\section{From $pp$ to $pA$ collisions}

Having fixed the free parameters in $pp$ collisions, we can proceed
and compute the absolute transverse spectra in $pA$ collisions.
We assume the proton and the deuteron to interact as
pointlike objects at an impact parameter $b$ with the nucleus.
Its nucleons, $N$, have isospin averaged parton distribution functions
$f_{i/N} = Z f_{i/p} + (A-Z) f_{i/n}$, with $A$ and $Z$ the atomic
mass and atomic number. 
Furthermore, we assume  that $A$-nucleus partons scatter only once on
the proton or the deuteron, due to their small density.
Then we may generalize \eqeqref{ppcoll_pt} as follows, 
without introducing further free parameters:
\begin{eqnarray}
  \frac{d\sigma}{d^2p_T dy d^2b}^{\hspace{-0.5cm}pA\rightarrow iX}
    \hspace{-0.5cm} &=& \Bigg\{
    \vev{xf_{i/p}}_{y_i,p_T} \, \frac{d\sigma^{\,iA}}{d^2p_T dy_i d^2b} 
    \In{y_i=y}
 \label{pAcoll} \\*
  && + T_A(b) \sum_{b}\, \vev{xf_{i/A}}_{y_i,p_T} \, 
    \frac{d\sigma^{\,ip}}{d^2p_T dy_i}  
    \In{y_i=-y} \Bigg\} \ .
     \nonumber
\end{eqnarray}
Hadron production is then computed analogously to \eqeqref{pptohadron}.

\begin{figure}[tb]
\begin{center}
\parbox[c]{7.5cm}{\epsfig{figure=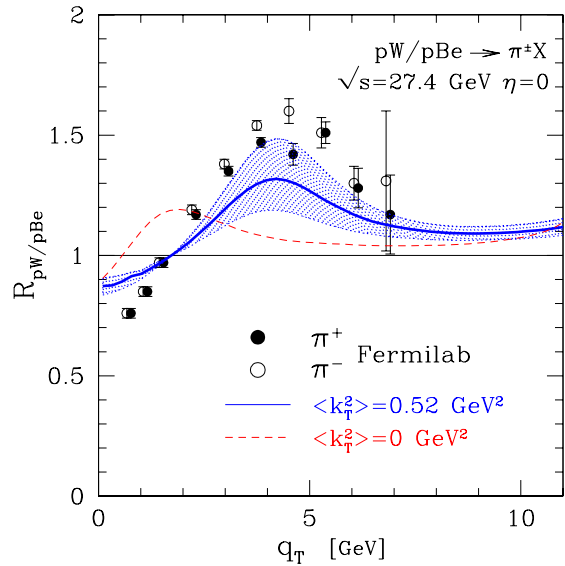,width=7.5cm}}
\vskip-.2cm
\caption{\footnotesize
Cronin effect in charged pion production at $\sqrt s=27.4$ GeV. 
Plotted is the ratio of the minimum bias charged pions $q_T$ spectrum 
at $\eta=0$ in $pW$ and $pBe$ collisions. The solid line is for a
scale choice $Q_p=Q_h=m_T/2$ and intrinsic $\vev{k_T}=0.52$ GeV$^2$.
The theoretical error due to the uncertainty in 
$p_0=0.8\pm0.1$ GeV is shown as a dotted band. 
The dashed line shows the result without intrinsic $k_T$ (the
theoretical uncertainty is not shown in this case). 
Data points taken from
\refref{Antreasyan79}.
}
\label{fig:FLAB.pi}
\end{center}
\vskip-.4cm
\end{figure}

Note that, due to \eqeqref{highptlim}, at large $p_T$, or as
$A\ra1$ [assuming $T_A(\vec b)\ra \delta(\vec b)$], 
the $b$-integrated $pA$ cross section reproduces exactly the $pp$
cross section discussed in the previous section. In this way, we can
calculate consistently both the $pp$ and $pA$ transverse spectra in
the same formalism. 

\begin{figure*}[tbh]
\begin{center}
\parbox[c]{15.5cm}{\epsfig{figure=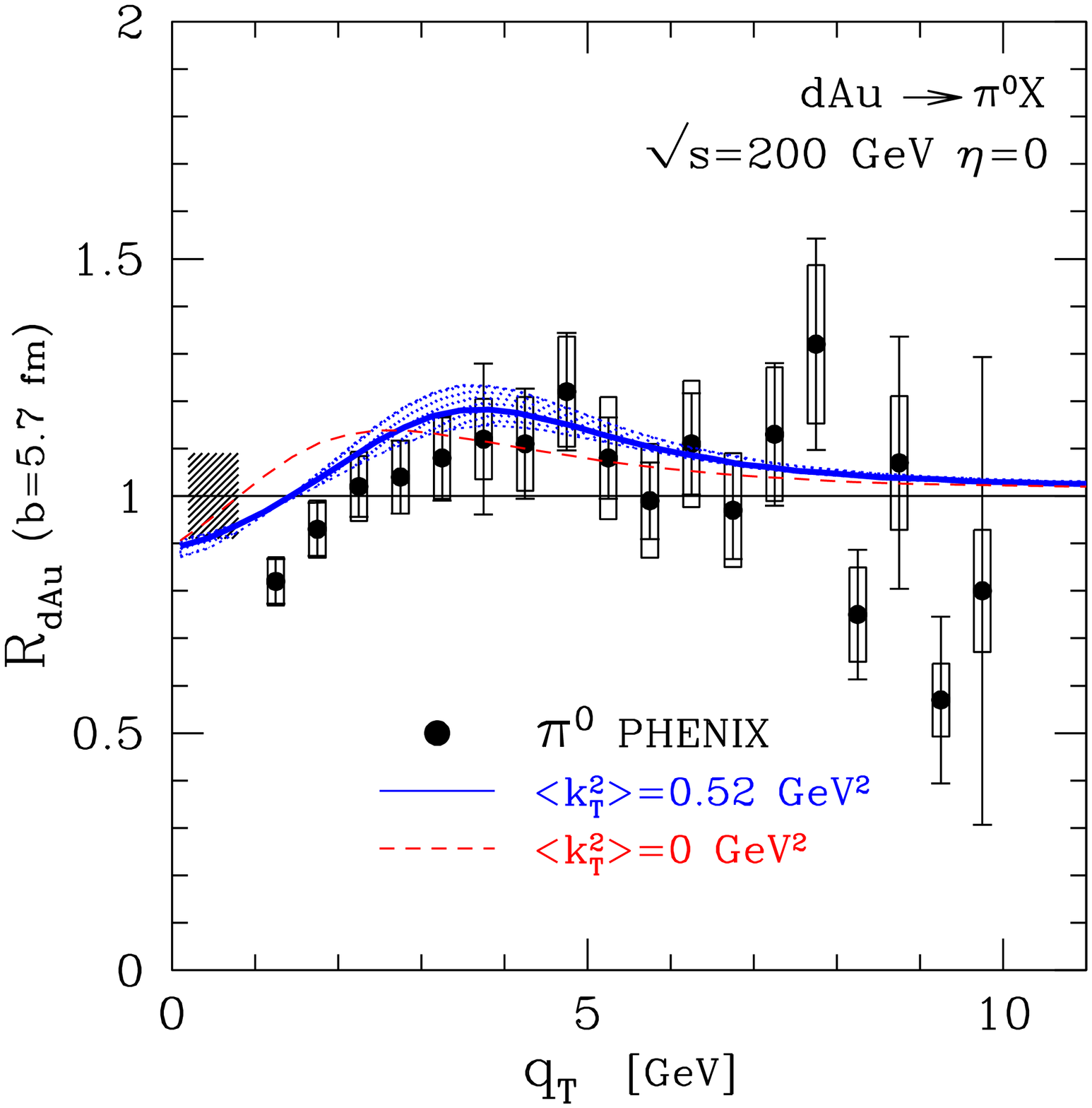,width=7.5cm}
               \ \ \epsfig{figure=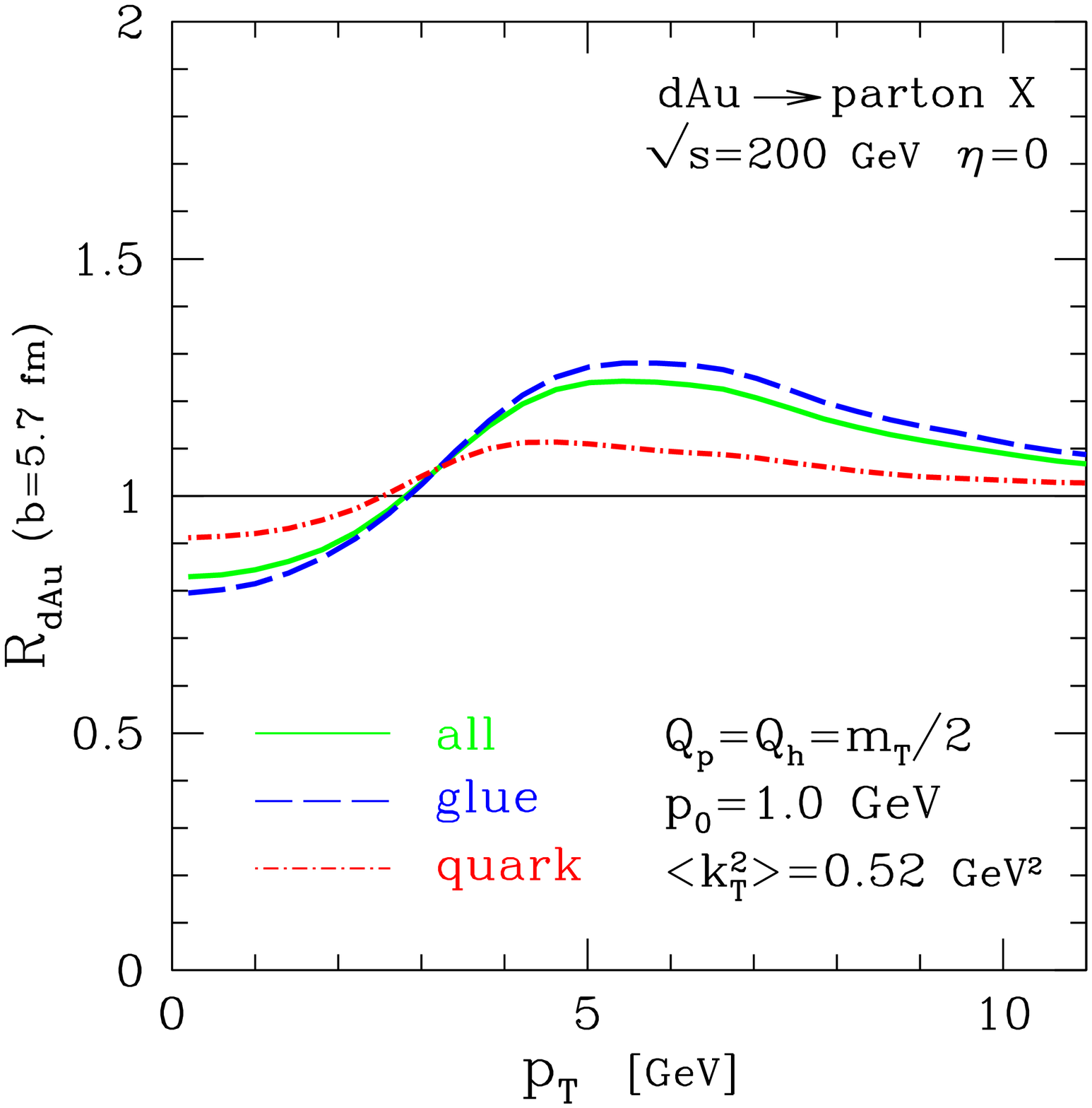,width=7.5cm}}
\vskip-.2cm
\caption{\footnotesize
Cronin ratio in minimum bias d+Au collisions at $\sqrt s=200$ GeV. 
{\it Left:} 
Cronin effect on neutral pion production. The solid line is for
$Q_p=Q_h=m_T/2$ and $\vev{k_T}=0.52$ GeV$^2$.
The theoretical error due to the uncertainty in 
$p_0=1.0\pm0.1$ GeV is shown as a dotted band.
The dashed line is computed with no
$k_T$ smearing, $\vev{k_T^2}=0$ GeV$^2$, and its theretical
uncertainty is not shown.
Data points are from the PHENIX collaboration, \refref{PHENIX}. 
Error bars represent statistical errors. The empty bands show
systematic errors which can vary with $q_T$. 
The bar at the left indicates the systematic uncertainty in the
absolute normalization of the pA cross section.
{\it Right:} Cronin Effect on gluon (dashed line), quark (dot-dashed) and
averaged quark and gluon production (solid), with $Q_p=Q_h=m_T/2$ and 
$\vev{k_T^2}=0.52$ GeV$^2$.
}
\label{fig:PHENIXpi0}
\end{center}
\vskip-.4cm
\end{figure*}

The Cronin ratio, $R_{BA}$, of the inclusive
differential cross sections for proton scattering on two different 
targets, normalized to the respective atomic numbers $A$ and $B$ is given by
\begin{equation}
    R_{BA}(q_T) = \frac{B}{A} 
        \frac{d\sigma_{pA}/d^2q_Tdy}{d\sigma_{pB}/d^2q_Tdy} \ .
 \label{Cronrat}
\end{equation}

First, we can test the GE formalism against low-energy data at $\sqrt
s=27.4$ GeV \cite{Antreasyan79} for the ratio of of midrapidity
$\eta=0$ pion spectra in
proton-tungsten ($pW$) and proton-berillium ($pBe$) collsions:
\begin{equation}
  R_{pW/pBe} \simeq 
    {\dfrac{d\sigma}{dy dq_T^2 d^2b}
       ^{\hspace{-0.5cm}pW\rightarrow \pi^\pm X}
       \hspace{-0.8cm} (b= b_W) } \Bigg/
    {\dfrac{d\sigma}{dy dq_T^2 d^2b}
       ^{\hspace{-0.5cm}pBe\rightarrow \pi^\pm X}
       \hspace{-0.8cm} (b=b_{Be})}
 \label{pWpBe}, 
\end{equation}
In our computations we approximated the minimum bias cross sections in
\eqeqref{Cronrat}, by computing \eqeqref{pAcoll} at an average impact
parameter $b_W=5.4$ fm and $b_{Be}=2.3$ fm, respectively. These values
were computed with the Monte Carlo model \cite{Miskowiecz}, in order for a
$pA$ collision at fixed impact parameter to produce the same number of
participant nucleons as a minimum bias one.
The result is shown in \figref{fig:FLAB.pi}. The two choices of scale,
$Q_p=Q_h=m_T/2$ and $Q_p=Q_h=m_T/2$ -- along with the respective fits
of $p_0$ and $K$ from Table~\ref{table:p0Kfits} -- 
give approximately the same result, and only the former choice is used in the
figure. The dotted area represents the theoretical error due to the
uncertainty in the fit of $p_0=0.7\pm 0.1$ GeV. The computation
reproduces satisfactorily the experimental data inside the theoretical
errors.   

Turning to $\pi^0$ production at $\sqrt s=200$ GeV at $|y|\leq0.3$, 
in the left panel of \figref{fig:PHENIXpi0} we compare our computation for 
\begin{equation}
  R_{dAu} \simeq 
    {\dfrac{d\sigma}{dq_T^2 dy d^2b}
       ^{\hspace{-0.5cm}dAu\rightarrow \pi^0 X}
       \hspace{-0.8cm} (b= b_{Au}) } \Bigg/
    T_{Au}(b_{Au}) {\dfrac{d\sigma}{dq_T^2 dy}
       ^{\hspace{-0.3cm}pp\rightarrow \pi^0 X}
       \hspace{-0.8cm} } \ ,
 \label{dAu}
\end{equation}
with $b_{Au}=5.7$ fm, 
to experimental data from the PHENIX collaboration \cite{PHENIX}.
The results obtained with the two choices of scales are similar, and
only the computation with $Q_p=Q_h=m_T/2$ is shown.
At this energy the sensitivity of the result to the error in the fit
of $p_0$ and to the scale choice is smaller than at Fermilab energy, 
thanks to the reduced steepness of the $pp$ spectrum. 
The result is compatible with data on the whole $p_T$ range 
inside the experimental statistical and systematic errors. 
Despite this caveats, the GE model tends to slightly overestimate the data 
at $p_T\lesssim 2$ GeV. What we see in \figref{fig:PHENIXpi0} is therefore a
possible indication for a dynamical shadowing in addition to the basic Glauber
geometrical shadowing. Its magnitude is consistent with the range
of dynamical shadowing explored in \cite{ktbroadening2,Vitev03,ktbroadening3} 
using a variety of shadowing functions \cite{shadfns}.  

In the right panel of \figref{fig:PHENIXpi0}, we plotted the
corresponding Cronin effect at the parton level. The Cronin ratio
peaks at fairly large $p_T\simeq 6$ GeV, compatible with the expected
$\vev{z}\simeq 0.6$.   
The peak in our computation is positioned at significantly larger transverse
momentum than 
found in the MV model of Ref. 
\refref{GJ03}. This difference may be due to their choice of parameters:
they compute the Cronin ratio $R_{BA}$ for a nucleus $A$ such that
$Q_{s/A}=2-3$ GeV, and an arbitrary  reference $B$ 
such that $Q_{s/B}=1$ GeV. Also an infrared cutoff $\Lambda=200$ MeV
was employed. The same value of the peak ($p_T\simeq 3$ GeV) was found
in the MV model via  numerical computation of \refref{JNV03}, with a
slightly different choice of parameters. 
None of the values of $Q_{s/A}$, $Q_{s/B}$ and $\Lambda$ where fixed
by fitting absolute inclusive spectra in $pp$ or $pA$ collision. 
Therefore, the results of \refref{GJ03} should be understood 
as illustrative of the qualitative features of the MV model. 
As \figref{fig:PHENIXpi0} clearly demonstrates, fragmentation strongly
distorts the features of the parton level Cronin effect. Therefore,
the transverse momentum scales illustrated  in the qualitative
saturation models as in \refsref{BKW03,KKT03,Albacete03,GJ03}, which
up to now do not attempt to include the distortions of scales due to
hadronization processes, should not yet be taken literally. 

\begin{figure}[bt]
\begin{center}
\parbox[c]{7.5cm}{\epsfig{figure=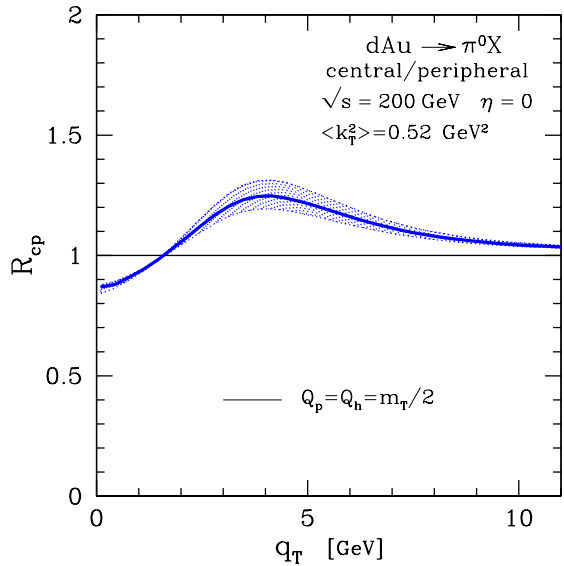,width=7.5cm}}
\vskip-.2cm
\caption{\footnotesize
Central to peripheral ratio for neutral pion production in minimum
bias d+Au collisions at $\sqrt s=200$ GeV. 
The central 0-20\% bin corresponds to an average
$b=3.5$ fm and the peripheral 60-88\% bin to an average $b=6.5$ fm.
The solid line is computed with a scale choice $Q_p=Q_h=m_T/2$ 
The theoretical error due to the uncertainty in 
$p_0=1.0\pm0.1$ GeV is shown as a dotted band.
}
\label{fig:PHENIXcp}
\end{center}
\end{figure}

To understand better the possible emergence of dynamical shadowing at
RHIC, we have two powerful handles. The first one is the rapidity
dependence of the Cronin effect, which we will
address in a separate publication (see also
\cite{Vitev03,GJ03,Albacete03} for a discussion in the framework of
approximated GE, MV and saturation models).
The second  handle is the centrality dependence of the Cronin
effect: the more central the collision, the higher the
density of target partons, the higher the shadowing effects induced by
non-linear parton interactions.
A very nice observable will be, in this repect, the ratio of $p_T$
spectra in central and peripheral collisions. This ratio has the additional
experimental advantage that most of the systematic errors shown in
\figref{fig:PHENIXpi0} are expected to cancel out. This will
provide  a rather precise
comparison to the GE model prediction which isolates geometric
shadowing only. Our predictions is shown in \figref{fig:PHENIXcp} for
two different choices of scale. The average impact parameter is 
$b=3.5$ fm and $b=6.5$ fm for central and peripheral
collisions, corresponding to centrality classes 0-20\% and 60-88\%,
respectively \cite{JacakINT}.
In the figure, the theoretical uncertainty due to the
uncertainty in the fit of $p_0 = 1.0 \pm 0.1$ GeV is shown as a dotted band.
If dynamical shadowing is present, we would
expect a larger deviation of the plotted curve from the data 
than what is observed in minimum bias collisions in
\figref{fig:PHENIXpi0}.

\section{Conclusions}

We have studied the Cronin effect in $pA$ and $dA$ collisions in the
context of Glauber-Eikonal models. These models incorporate parton
multiple scatterings and unitarity in pQCD in a consistent way. 
Moreover, they include a detailed parton kinematics and reproduce, in
the limit of $A=1$, the hadron transverse spectra as computed in the 
pQCD parton model. The analysis of $pp$ spectra allows to fix the
free parameter of the model, and to compute the spectra in $pA$
collisions without further assumptions. 

A powerful feature of GE models is that they automatically include in
the computation geometrical shadowing effects induced by unitarity and
parton multiple scatterings. By isolating geometrical shadowing,
one can use the GE model computations as a baseline in the search for genuine  dynamical shadowing effects due to non-linear
parton interactions.

We tested our computation of the Cronin effect in minimum bias collisions 
at $\sqrt s = 27.4$ GeV. The same formalism applied to the recently
measured $d+Au$ data at $\sqrt s=200$ AGeV
\cite{PHENIX,STAR,PHOBOS,BRAHMS} describes well the Cronin effect at
large $p_T$, with a tendence to overestimate by $\sim 10-20\%$ the effect for 
$\pi^0$ at $\eta=0$ and $p_T\lesssim 2$ GeV. Our results are surprisingly
similar to predictions based on  phenomenological approximated GE models 
\cite{ktbroadening1,ktbroadening2,Vitev03,ktbroadening3} 
in spite of the inclusion of geometrical shadowing in our GE approach. 
This provides further evidence
for the possible existence of moderate shadowing in the $x\sim 0.01-0.1$
range as explored in those references. 
However, radical gluon shadowing as predicted in \cite{CGCandsuppression}   
is not supported by the data. It remains to be seen
if the most recent  variations of saturation models can be fine tuned
to account 
to the thusfar featureless $R_{dAu}\sim 1$ RHIC data.
Future analysis of the centrality and pseudo-rapidity
dependence of the Cronin effect at RHIC will provide a powerful tool to
further constrain  the magnitude of the dynamical shadowing effect.

\begin{acknowledgments}
\vskip.3cm
We are grateful to B.Jacak, J.Jalilian-Marian,
B.Kopeliovich, Y.Kovchegov, H.Kowalski, P.Levai, A.Metz, N.N.Nikolaev,
R.Venugopalan, I.Vitev and the participants of the ``Columbia Nuclear
Theory Visitor Program 2003'' for 
inspiring discussions. A special thank to H.Honkanen for her help
with computations of $pp$ collisions. A.A. thanks the Institute of
Nuclear Theory at the University of Washington for its hospitality,
during which part of this work was done.

\end{acknowledgments}


\end{document}